\documentclass[aps, physrev,twocolumn,preprintnumbers,superscriptaddress,longbibliography,showkeys,amsmath,amssymb]{revtex4-2}
\usepackage{graphicx}% Include figure files
\usepackage{dcolumn}% Align table columns on decimal point
\usepackage{bm}% bold math
\usepackage[mathlines]{lineno}% Enable numbering of text and display math
\nolinenumbers % Commence numbering lines
\usepackage{easybmat}
\usepackage{ragged2e}
\usepackage{amstext}
\usepackage{titlesec}
\usepackage{color}
\usepackage{braket}
\usepackage{soul}
\usepackage{subcaption}
\usepackage{float}
\usepackage{caption}
\usepackage{appendix}
\usepackage{natbib}
\usepackage{calligra,lmodern}
\usepackage[colorlinks=true,citecolor=blue,urlcolor=blue, linkcolor= magenta]{hyperref}
\usepackage{footnote}
\usepackage[%showframe,%Uncomment any one of the following lines to test 
scale=1,% marginratio={1:1, 2:3}, %ignoreall,% default settings
text={7in,10in},centering,
height=10in,a4paper,%hmargin={3cm,0.8in},
]{geometry}
\newcommand{\be}{\begin{equation}}
	\newcommand{\ee}{\end{equation}}
\newcommand{\ba}{\begin{eqnarray}}
	\newcommand{\ea}{\end{eqnarray}}

\usepackage{tikz,xcolor,hyperref}

\definecolor{lime}{HTML}{A6CE39}
\DeclareRobustCommand{\orcidicon}{\hspace{-4pt}
	\begin{tikzpicture}
		\draw[lime, fill=lime] (0,0) 
		circle [radius=0.16] 
		node[white] {\hspace{0.1mm}{\fontfamily{qag}\selectfont \tiny ID}};
		\draw[white, fill=white] (-0.07,0.1) 
		circle [radius=0.01];
	\end{tikzpicture}
	\hspace{-3.2mm}
}

\foreach \x in {A, ..., Z}{\expandafter\xdef\csname 
	orcid\x\endcsname{\noexpand\href{https://orcid.org/\csname orcidauthor\x\endcsname}
		{\noexpand\orcidicon}}
}

\begin{document}

%\preprint{APS/123-QED}

\title{Lepton asymmetry leading to baryogenesis by primordial black holes}

 \author{Mriganka Dutta\orcidA{}}\email{mrigankad@iisc.ac.in}
\affiliation{Department of Physics, Indian Institute of Science, Bangalore 560012, India}
  \author{Banibrata Mukhopadhyay\orcidB{}}\email{bm@iisc.ac.in}
\affiliation{Department of Physics, Indian Institute of Science, Bangalore 560012, India}
\affiliation{Joint Astronomy Programme, Department of Physics, Indian Institute of Science, Bangalore 560012, India}
  \author{Abhishek Kumar Jha\orcidC{}}\email{kjabhishek@iisc.ac.in}

  \affiliation{Department of Physics, Indian Institute of Science, Bangalore 560012, India}
  \author{Mayank Pathak\orcidD{}}\email{mayankpathak@iisc.ac.in}
  \affiliation{Joint Astronomy Programme, Department of Physics, Indian Institute of Science, Bangalore 560012, India}
  \author{Siba Prasad Das\orcidE{}}\email{spd.phy@unishivaji.ac.in}
  \affiliation{Department of Physics, Shivaji University, Kolhapur 416004, India}

\begin{abstract}
Baryogenesis remains an unresolved problem in cosmology, with existing mechanisms facing significant caveats. We show that the effects of primordial black holes (PBHs) on neutrinos  produce the lepton asymmetry $\sim 10^{-10}$ which subsequently produces the baryon asymmetry. We consider the Dirac Lagrangian in curved spacetime in local coordinates exhibiting Hermitian pseudo-vector and non-Hermitian vector terms. These terms lead to energy splitting between weakly interacting neutrinos and antineutrinos, resulting in their unequal number densities and hence a lepton asymmetry. While the non-Hermitian effect leads to a non-conserved total probability of neutrinos, the leptogenesis due to gravitational effects of a PBH could be significant until the nucleosynthesis era. This in turn produces baryon asymmetry from the symmetry of lepton and baryon numbers via the sphaleron process in the electro-weak era. We show that in the most conservative scenario, the PBHs of mass $\sim 10^{12}$ g and spin $\sim 0.01$ produce the observed baryogenesis at temperature 130 GeV, when such PBHs are available abundantly. However, massive PBHs also could produce the observed asymmetry, assuming the non/anti-Hermitian vector couplings for neutrino and anti-neutrino get canceled from the Lagrangian, leading the system to be Hermitian.

 \end{abstract}

\maketitle
\section{Introduction}

One of the long standing problems in the BigBang (BB) cosmology is the prediction of 
asymmetry of baryon number ($B$), namely $\Delta B$, as we observe. More precisely, the difference in the number densities between baryon ($n_B$) and antibaryon ($n_{\bar B}$) per photon number density ($n_\gamma$): $(n_B-n_{\bar B})/n_\gamma\sim 10^{-10}$, set in the early universe. There are many explorations to explain $\Delta B$ (e.g. \cite{Sakharov:1967dj,Fukugita:1986hr,Affleck:1984fy,Riotto:1998bt}, to mention a few), however, none of them is beyond any caveats. Often, $\Delta B$ is argued to originate from the lepton number ($L$) asymmetry $\Delta L$ by the $B-L$ symmetry and the sphaleron process. In this work, we show that neutrinos being lepton get asymmetrized in the presence of strong gravity of black holes (BHs), which leads to baryon asymmetry.  

Neutrinos being spinors, we start with
the Dirac equation in curved spacetime as
\begin{equation}
    (i\gamma^\mu\nabla_\mu+m)\Psi=0,
    \label{1}
\end{equation}
where $\gamma^\mu$ are Dirac $\gamma$-matrices, $\nabla_\mu$ is the covariant derivative including spin-connections, and $\Psi$ represents the spinor. 
We adopt natural units by setting $G M= c = \hbar = 1$, where $G$ is Newton's gravitation constant, $\hbar$ is the reduced Planck constant, $c$ is the speed of light, and $M$ is the mass of the BH. Distances are measured in the units of BH’s gravitational radius $r_g=GM/c^2$ ($=1$ here).

The spin-curvature connections for neutrinos and anti-neutrinos lead to split in their single-particle energies assuming their plane wave nature in local inertial coordinates. This gravity induced process with lepton number violation further creates nonzero $\Delta L$ from the statistical process, which leads to $\Delta B$.

It is well-known that the Dirac Hamiltonian in curved spacetime leads to one vector term and another axial-vector term (for that the spacetime can not be static but may be stationary, like the Kerr metric), particularly in the local inertial coordinates  \cite{Mohanty:2002et,Debnath:2005wk,Mukhopadhyay:2007vca,Sinha:2007uh,Huang:2008kh}. Interestingly, the vector term could be non-Hermitian (see, e.g., \cite{Huang:2008kh}), even anti-Hermitian, an indicative of local dissipation due to spin-curvature interaction. The axial vector term splits the energies between up and down spins. This has many consequences, including those related to quantum correlation, speed-limit, gravitational geometric phases \cite{  Mukhopadhyay:2018oli, Dixit:2019lsg, Ghosh:2020lhw, Jha:2024gdq, Jha:2024asl, Jha:2025ekn}. 

The standard model neutrinos are left-handed and anti-neutrinos are right-handed. Therefore, the above mentioned axial vector essentially splits neutrino--anti-neutrino energies in local coordinates in the presence of background gravitational fields. Neutrinos are in many astrophysical and cosmological sites, e.g. core of neutron stars, early universe, supernova, which are often involved with strong gravity
(see, e.g., \cite{Chen:2006rra,Mukhopadhyay:1999ut,Banerjee:2013uba}). Naturally, near a compact object, where the effect of curvature is maximum on this leptonic system (in principle for any fermion), the energy difference between neutrinos and anti-neutrinos is maximum \cite{Jha:2025ekn}, 
which diminishes with increasing distance. 
The validation of local flat spacetime allows us to write a Fermi-Dirac distribution of this spin-1/2 system and thus estimation of the number density.

In this work, we explore the solution of Eq.\,(\ref{1})
in the background of BH gravity, particularly the Kerr metric, and apply it for a neutrino--anti-neutrino system. We will show that primordial BHs (PBHs) can 
produce the required neutrino asymmetry leading to the observed $\Delta B$ via sphaleron process by the era of electro-weak phase transition of temperature $T=130$ GeV \cite{Riotto:1998bt}. 
According to the standard model, the rate of sphaleron process at a temperature $T$ is $\Gamma(T)\approx \kappa T^4 \exp(-E_{\rm sph}(T)/T)$ \cite{Rubakov:1996vz}, where $E_{\rm sph}$ is the energy of sphaleron and $\kappa$ is a constant. As the universe cools, the surrounding temperature decreases. Thus, sphaleron rate decreases and for $T< 130$ GeV this rate becomes negligible enough to not contribute.
Subsequently, after the BB nucleosynthesis (BBN) era, the neutrinos are expected to be out of equilibrium. Hence, $\Delta B$ set by the electro-weak era will sustain throughout. Therefore, the conditions, equivalent to those of Sakharov \cite{Sakharov:1967dj}, strictly to follow for genesis are: (1) the BH has to be rotating leading to the deviation of the spacetime from static to stationary, (2) weakly interacting neutrinos and anti-neutrinos to acquire difference in their effective mass (i.e. dispersion energy) leading to neutrino asymmetry by simple statistical process, (3) sphaleron to produce baryon asymmetry which is to be out of equilibrium so that the created asymmetry sustains. 

The assumptions we consider are: (i) the underlying gravitational field acting on the quantum system is classical and there is no back reaction of the spinor onto the background, (ii) a locally inertial frame is possible to choose for the problem, (iii) the characteristic length over which the (pseudo)vector potentials change is more than the wavelength of the neutrinos 
and, hence, a plane wave solution for the spinor can be taken locally, and (iv) since we can define a local solution, a definite energy of the neutrino--anti-neutrino system can be derived which further leads to the Fermi-Dirac statistical system.

\section{Number Densities for Neutrino and Anti-neutrino}

The covariant derivative with spin-connections of Eq.\,(\ref{1}) can be expanded in local coordinates as \cite{Schwinger:2019lox, Parker:1980hlc}
\begin{equation}
    \gamma^\mu\nabla_\mu\Psi=\gamma^a\nabla_a\Psi=\gamma^a(\partial_a -\frac{i}{4}\omega_{bca}\sigma^{bc})\Psi,
    \label{2}
\end{equation}
where Greek and Latin indices, respectively, correspond to global curved and local flat coordinates, both running for one temporal and three spatial coordinates. After expanding the 
spin-connections, we obtain the Dirac equation as \cite{ Mohanty:2002et, Dixit:2019lsg, Parker:1980kw}
\begin{equation}
     (i\gamma^{a}\partial_{a} - m)\Psi+(iA_a\gamma^a+B^a\gamma_a\gamma^5)\Psi=0.
     \label{3}
\end{equation}
The axial vector term $\gamma_aB^a\gamma^5$ vanishes for a Schwarzschild BH, when 
\begin{equation}
   A_a=(\omega_{bac}-\omega_{abc})\eta^{bc} \hspace{0.1 cm} \text{and}    
\hspace{0.1 cm}    B^a=\omega_{bcd}\epsilon^{adbc},
\label{4}
\end{equation}
and the spin connections are expressed by vierbiens $e^\mu_a$ as
\begin{equation}
    \omega_{bca} = e_{b\lambda} (\partial_a e^{\lambda}_{c} + \Gamma^{\lambda}_{\alpha\mu}e^{\alpha}_{c}e^{\mu}_{a}),~{\rm with}~g^{\mu\nu}=e^\mu_a e^\nu_b\eta^{ab},
    \label{5}
\end{equation}
whereas $g^{\mu\nu}$ and $\eta^{ab}$ are curved and flat spacetime matrices respectively.

\begin{figure*}[!htbp]
  \centering
  \begin{subfigure}[b]{0.35\textwidth}
    \centering
    \includegraphics[width=\textwidth]{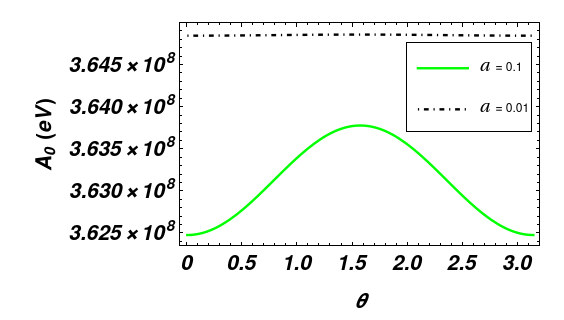}
    \caption{}
    \label{1_fig:suba}
  \end{subfigure}
 % \hfill
  \begin{subfigure}[b]{0.3\textwidth}
    \centering
    \includegraphics[width=\textwidth]{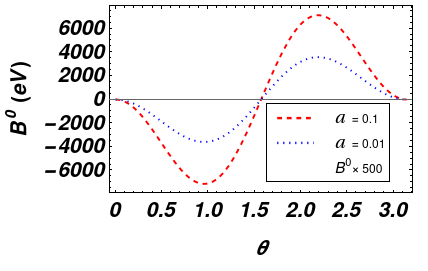}
    \caption{}
    \label{1_fig:subb}
  \end{subfigure}
  \begin{subfigure}[b]{0.3\textwidth}
    \centering
    \includegraphics[width=\textwidth]{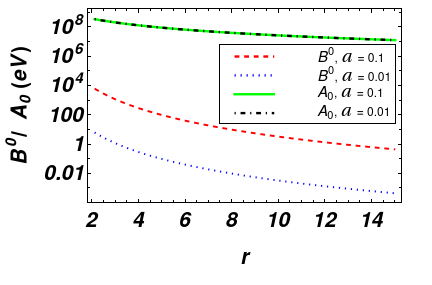}
    \caption{}
    \label{1_fig:subc}
  \end{subfigure}
  \caption{\justifying{{The variation of $A_0$ and $B^0$ as functions of (a)-(b) coordinate $\theta$ of BH spacetime at $r=2.1$, and (c) distance from BH for $\theta=\pi/4$, where BH mass $M=10^{-18}M_{\odot}$. In Fig.\,\ref{1_fig:subb}, $B^0$ for $a=0.01$ is multiplied by 500 to keep in the same scale as of $a=0.1$ in the plot.}}}
  \label{fig1}
\end{figure*}
Also the vector term is effectively non-Hermitian (see \cite{Huang:2008kh} for details). In short, the background gravitational effects, particularly in local coordinates, effectively lead to dissipation.

\subsection{Energy dispersion relation}

Now assuming a local plane wave solution for $\Psi$ in Eq.\,(\ref{3}), the energies of the spin-up and spin-down particles, which are respectively of neutrino ($\nu$) and anti-neutrino ($\bar{\nu}$), come out to be
\begin{align}
    &E_{\nu,\bar{\nu}}=\sqrt{(\vec{p}\mp\vec{B}+i\vec{A})^2+m^2}\pm B^0-iA_0, 
    \label{6}
\end{align}
where $B^0$ and $A_0$ are the temporal components of $B^a$ and $A_a$, respectively.
Assuming neutrinos and anti-neutrinos have high kinetic energy (i.e. momentum) compared to other energy terms, Eq.\,(\ref{6}) can be approximated to
\begin{align}
&E_{\nu,\bar{\nu}}^R=|\vec{p}|+\frac{m^2+|\vec{B}|^2-|\vec{A}|^2}{2|\vec{p}|}\pm B^0,\hspace{0.5cm}
E_{\nu\bar{\nu}}^I=-A_0,
\label{8}
\end{align}
where the superscripts $R$ and $I$ indicate the real and imaginary components, respectively. Here we omit 
$\vec{A}\cdot\vec{B}$, $\vec{p}\cdot\vec{A}$ and $\vec{p}\cdot\vec{B}$ by taking average over the angle between those vectors as particles can propagate from all possible directions. Again, to make calculation easier, we terminate the second term in $E_{\nu\bar{\nu}}^R$. This signifies that the change of energy due to $B^a$ and $A_a$ will be less than the energy corresponding to the temperature of the surroundings.

\subsection{Number density}

Now, statistically, we obtain the number density difference (hence asymmetry) between weakly interacting neutrino and anti-neutrino as 
\begin{equation}
    \Delta n=\frac{g}{(2\pi)^3}\int \mathrm{d}^3p \left[ \frac{1}{1+e^{{E_{\nu}}/T}}-\frac{1}{1+e^{{E_{\bar{\nu}}}/T}}\right].
    \label{9}
\end{equation}
Clearly $\Delta n\neq 0$ as $E_\nu\neq E_{\bar\nu}$.
Hence, the asymmetry is given by
\begin{align}
\Delta n_R \approx \frac{g}{(2\pi)^3} \left(\frac{1}{3}B^0\pi^2T^2+\frac{{B^0}^3}{3}-A_0^2B^0\right),\nonumber\\
    \Delta n_I \approx \frac{g}{(2\pi)^3} \left(4A_0B^0T\log(2)+\frac{A_0{B^0}^3}{6T}-\frac{A_0^3B^0}{6T}\right).
    \label{10}
\end{align}

Clearly from Eq.\,(\ref{10}), asymmetry survives only for $B^0\neq 0$.
The average asymmetry of number density per photon over the entire region around a BH leads to 
 \begin{equation}
     \left|\frac{\Delta n}{n_\gamma}\right|_{avg}=\frac{\int |\Delta n| d\theta}{\int |n_\gamma| d\theta},
     \label{12}
 \end{equation}
 where $|\Delta n|=\sqrt{(\Delta n_R)^2+(\Delta n_I)^2} $ and $n_\gamma$ is the photon number density.

The quantities $A_a$ and $B^a$ are highly dependent on the distance we consider from the BH. We shall see that far away from the BH, the effect of gravity diminishes rapidly, and thus $|\Delta n|$ decreases. Hence, in our scenario we shall consider regions very near to BH's horizon, which will contribute the most to $|\Delta n|$. Also, the asymmetry stabilizes when neutrinos go out of equilibrium beyond certain expansion.

\section{Neutrino--anti-neutrino transition probability}

In the Weyl representation, a neutrino doublet can be expressed as $\Psi=\begin{pmatrix}
    \psi^c\\
    \psi
\end{pmatrix}$,
where $\psi^c$ and $\psi$ are two-component spinors corresponding to the anti-neutrino and neutrino, respectively. These are the eigenstates of the lepton number operator with eigenvalues -1 and +1 respectively. Following Eq.\,(\ref{3}) and the previous work \cite{Sinha:2007uh}, the effective mass matrix of the neutrino--anti-neutrino system under gravity can be cast to
\begin{equation}
    \mathcal{M}=\begin{pmatrix}
    -B^0-iA_0& -m\\
    -m &  B^0-iA_0\\
  \end{pmatrix},
  \label{13}
\end{equation}
where $m$ is the Majorana mass. Clearly, this effective mass matrix is non-Hermitian ($\mathcal{M}^\dagger \neq \mathcal{M}$), however can be diagonalized by a unitary transformation with
\begin{eqnarray}
U(\theta_\nu)=\begin{pmatrix}
    \cos\theta_\nu & \sin\theta_\nu\\
    -\sin\theta_\nu & \cos\theta_\nu
\end{pmatrix}.
\end{eqnarray}
Now, following previous works \cite{Barenboim:2002hx, Sinha:2007uh,Jha:2025ekn}, we propose
neutrino and anti-neutrino states are known with their definite
spin and their unitary transformed states, $\nu_1$ and $\nu_2$, are known with 
definite masses, in the same way of neutral kaon, so that
at time $t=0$, 
\begin{equation}
    \begin{pmatrix}
        \ket{\nu_1}\\
        \ket{\nu_2}
    \end{pmatrix}=U(\theta_\nu)\begin{pmatrix}
        \ket{\psi^c}\\
        \ket{\psi}
    \end{pmatrix}
    \label{14}
\end{equation}

with the mixing angle
\begin{equation}
    \theta_\nu=\tan^{-1}\left[\frac{m}{B^0+\sqrt{(B^0)^2+m^2}}\right].
    \label{15}
\end{equation}
Therefore, there is a probability of oscillation between 
$\nu_1$ and $\nu_2$, as $\psi$ and $\psi^c$ evolve with their
definite respective spin.

Due to distinct spin, hence energy, of the states $\ket{\psi^c}$ and $\ket{\psi}$, in the presence of the gravitational potentials $B^0$ and $A_0$, as given in Eq.\,(\ref{6}), their time evolutions are given by 
\begin{equation}
\ket{\psi^c(t)}\rightarrow\ket{\psi^c(0)}e^{-iE_{\bar{\nu}} t},\hspace{0.5cm}\ket{\psi(t)}\rightarrow\ket{\psi(0)}e^{-iE_{\nu}t}.
\label{16}
\end{equation}

In the ultra-relativistic regime, 
the oscillation probability at time $t$ is given by
\begin{equation}
    \mathcal{P}_{\text{osc}}(t)=\left|\left\langle\left. \nu_2(t) \right| \nu_1(0) \right\rangle \right|^2=[\sin^2(2\theta_\nu) \sin^2\left(B_0 t\right)]e^{-2A_0 t}.
    \label{17}
\end{equation}
Similarly, the survival probability is given by
\begin{eqnarray}
    &&\mathcal{P}_{\text{sur}}(t)=\left|\left\langle\left. \nu_1(t) \right| \nu_1(0) \right\rangle \right|^2\\ \nonumber 
    &&=[1-\sin^2(2\theta_\nu) \sin^2\left(B_0 t\right)]e^{-2A_0 t}.
    \label{18}
\end{eqnarray}

\begin{figure*}[!htbp]
  \centering
  \begin{subfigure}[b]{0.45\textwidth}
    \centering
    \includegraphics[width=\textwidth]{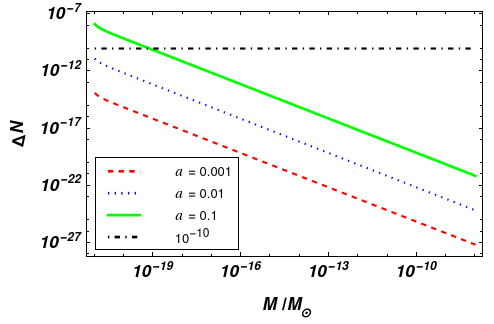}
    \caption{}
    \label{fig3:suba}
  \end{subfigure}
  \hfill
  \begin{subfigure}[b]{0.45\textwidth}
    \centering
    \includegraphics[width=\textwidth]{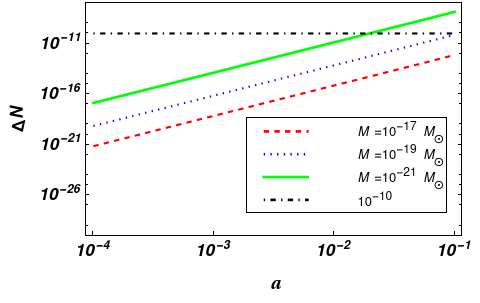}
    \caption{}
    \label{fig3:subb}
  \end{subfigure}
  \caption{\justifying{
  Variation of neutrino asymmetry as a function of (a) BH mass for various spin, and (b) BH spin for various mass, at $r=2.1$ and $T=130~ \mathrm{GeV}$. The dot-dashed horizontal line indicates $10^{-10}$ level.
  }}
  \label{fig3}
\end{figure*}

From Eq.\,(\ref{6}), we notice the presence of the imaginary term ($iA_0$) in the energy expressions, which leads to a decay of the state $\ket{\nu_{1,2}(t)}$ with a lifetime $(2A_0)^{-1}$. Consequently, the total transition probability
\begin{equation}
\mathcal{P}_{\text{tot}}(t) = \mathcal{P}_{\text{osc}}(t) + \mathcal{P}_{\text{sur}}(t) = e^{-2A_0 t}
\label{20}
\end{equation}
is not conserved until $A_0\rightarrow 0$ (at large $r$), indicating dissipation during the neutrino--anti-neutrino transition. 
Thus, due to the presence of non-Hermitian term in the Dirac equation in curved spacetime, the neutrino number asymmetry per photon, from Eqs.\,(\ref{6}), (\ref{9}) and (\ref{12}), is then explicitly given by
 \begin{equation}
\Delta N= \frac{\int |\Delta ne^{-2A_0 t}| \mathrm{d}\theta}{\int |n_{\gamma}| \mathrm{d}\theta}.
 \label{21}
\end{equation}

However, if one considers the total Lagrangian of the system including the neutrino and anti-neutrino states together, then vector (locally anti-Hermitian) terms are canceled out \cite{Sinha:2007uh,Schwinger:2019lox}. This recovers the conservation of the oscillation probability and the decay term $\exp[-2A_0t]$ in Eq.\,(\ref{21}) does not appear. 

\section{Asymmetry around a black hole}
For the present purpose, we consider a rotating black hole spacetime and choose the Kerr-Schild polar coordinate system, whose line element is given by \cite{Takahashi:2007yi}
\begin{align}
    ds^2&=-(1-2Mr/\rho^2)\mathrm{d}t^2+4Mr/\rho^2\mathrm{d}r\mathrm{d}t\nonumber\\&-(4Mar\sin^2\theta/\rho^2)\mathrm{d}t\mathrm{d}\phi+(1+2Mr/\rho^2)\mathrm{d}r^2\nonumber\\&+\rho^2 \mathrm{d}\theta^2-2a\sin^2\theta(1+2Mr/\rho^2)\mathrm{d}r\mathrm{d}\phi\nonumber\\&+A\sin^2\theta/\rho^2\mathrm{d}\phi^2,
    \label{7}
\end{align}
where $\rho^2=r^2+a^2 \cos^2\theta$, $A=(r^2+a^2)^2-a^2\Delta\sin^2\theta$, $\Delta=r^2-2Mr+a^2$, $a$ is the dimensionless specific angular momentum per unit mass ($-1\leq a\leq 1$) of BH, and $\theta\in [0,\pi]$ is the angle of the position vector of the spinor with
respect to the spin axis of the BH.

\subsection{Potentials in Kerr metric}
From Eqs.\,(\ref{4}) and (\ref{5}), one can calculate $B^a$ and $A_a$ for Eq.\,(\ref{7}). Fig.\,\ref{fig1} illustrates the variations of gravitational scalar potentials $B^0$ and $A_0$ as functions of $\theta$ and $r$ for spinning BHs of mass $M = 10^{-18} M_\odot$ ($=2\times 10^{15}$ g). As expected, potentials decrease with distance, however they are 
symmetric with respect to $\theta$, though the variation of $A_0$ with $\theta$ is very small. Moreover, the magnitude of $A_0$ is more than $B^0$, when increasing spin increases $B^0$ but not $A_0$ near the horizon.

\section{Neutrino Asymmetry}
We know that Eq.\,(\ref{9}) is applicable only if the neutrino--anti-neutrino system is in thermal equilibrium with surrounding. In the early universe, neutrinos were in thermal equilibrium when the interaction rate (via weak force) is faster than the expansion rate of  surrounding (universe). This is only possible till BBN era, i.e. before 1 s history of the universe when temperature is above 1 MeV \cite{Follin:2015hya}. Subsequently, the neutrino decoupling happens and neutrinos go
 out of equilibrium settling $\Delta N$. 
 The observed baryon asymmetry of $10^{-10}$, arising from $\Delta N$ via sphaleron process for $T\gtrsim 130$ GeV, can be obtained by exploring ranges of the mass and spin of PBHs, as is demonstrated below. For $T<130$ GeV, sphaleron process gets suppressed, hence no conversion from the lepton asymmetry to baryon asymmetry occurs.

\subsection{Primordial black hole}
Most of the PBHs are expected to form in the (very) early universe. During radiation dominated era, PBHs can grow by absorbing radiation \cite{DeLuca:2020bjf}. However, that era is expected to be highly isotropic, hence the PBHs are not expected to acquire angular momentum leading their effective spin to be very close to zero.
While in the matter dominated era, these PBHs can spin up by absorbing matter, their spin remains very small \cite{Mirbabayi:2019uph}. 
 
Whether a BH spins or not, $A_a$ survives. As the scalar potential $A_0$ is inversely proportional to the mass of BH, the decaying factor $\exp[-2A_0t]$ tends to unity as the mass of BH increases. Therefore, the lepton asymmetry does not change due to the non-Hermitian effects for BHs of mass comparable to solar or higher masses. However, for a lower mass PBH, $A_0$ is significantly larger and decreases the total transition probability affecting neutrino number density. Furthermore, as because $A_0$ is highest near the BH, as shown by Fig.\,\ref{1_fig:subc}, its decaying effect on the number density near a PBH also increases significantly. On the other hand, due to increasing $B^0$, on approaching a PBH the neutrino--anti-neutrino number density asymmetry increases due to increasing energy difference, as understood from Eqs.\,(\ref{6}) and (\ref{7}). Hence, the effects of
 $A_0$ and $B^0$ work oppositely on the asymmetry. 

 The time $t$ involved in the probability decay is determined by the timescale of interaction of the neutrino with the BH event horizon. This interaction is viable, as long as $r_g/c<1~{\rm s}$ (which is of the order of the horizon crossing timescale), which is quite valid for a wide range of BH mass, including all subsolar PBHs. 
The initial mass of a PBH, however, scales with the age of universe, given by $M\simeq 10^{15}~{\rm gm}~(t/10^{-23}~{\rm s})$ \cite{Carr:2020gox}. 

 For a PBH of $M=10^{14}$ g and $a=0.1$, 
 from Eq.\,(\ref{21}) we obtain $\Delta N\sim 10^{-10}$ for $T=130$ GeV near the horizon. This value further scales inversely with the BH mass and directly with BH spin. Fig.\,\ref{fig3} shows how $\Delta N$ changes with different masses and spins of BHs. 

 The figure confirms that the asymmetry inversely and directly scales with, respectively, the mass and spin of the BH, corroborating with the respective behaviors of $A_0$ and $B^0$ discussed above. It shows that $\Delta N\sim 10^{-10}$ for $M\lesssim 10^{-18}M_\odot$ and $a\lesssim 0.1$. Both appear to be viable PBH parameters. Formation of $\Delta N$ with $T\gtrsim 130$ GeV leads to the formation of (equal) baryon asymmetry by sphaleron process.

\subsection{Astrophysical black holes}
 Astrophysical BHs are very massive compared to PBHs, hence they, with their much larger size, will have a very small quantum effect on neutrinos. Since neutrinos are ultra-relativistic particles, the largest wavelength they can have is the Compton wavelength. This means that the quantum effect should start dominating from BHs with gravitational radius around that Compton wavelength, which corresponds to  $M\sim 10^{-8}M_\odot$.
 
 Stellar mass BHs generally have the mass range $10-100M_\odot$. 
 A $10M_\odot$ stellar mass BH with $a=0.5$ gives lepton asymmetry, in addition to the primordial one, of $10^{-19}$, which is too tiny to contribute.

\begin{figure}
  \centering
  \includegraphics[width=0.28\textwidth]{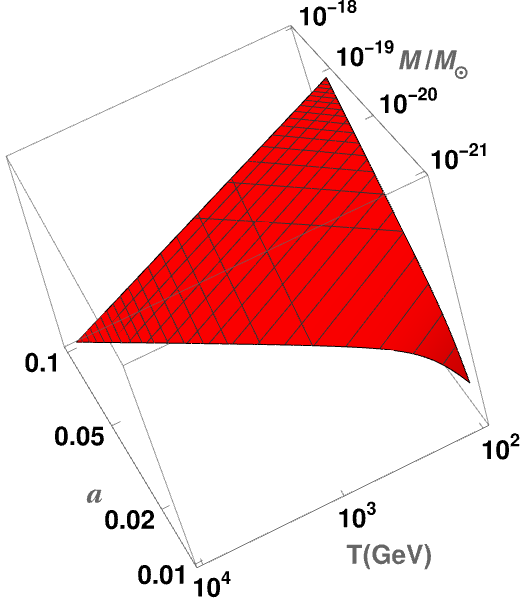}
  \caption{Three dimensional surface of $\Delta N=10^{-10}$ at $r=2.1$ in the ranges of BH mass, spin, and early universe temperature.}
  \label{3D}
\end{figure}

For supermassive BHs (SMBH), this asymmetry is even smaller. Considering the nearest SMBH to us, Sgr A$^*$, with inferred $M=4\times10^{6} M_\odot$ and $a=0.1$ \cite{EventHorizonTelescope:2022wkp,Fragione:2020khu,Broderick:2016ewk}, $\Delta N\sim 10^{-26}$. A SMBH of similar mass but high spin at most can lead to an asymmetry of $10^{-23}$. 

Most of the other SMBHs are massive, leading to much less additional asymmetry.  

\subsection{Viable parameter regime}
Given the observed asymmetry, we show in Fig.\,\ref{3D} the ranges of PBH mass and spin producing a surface of $10^{-10}$ for $10^2\lesssim T/{\rm GeV}\lesssim 10^4$. It suggests that at the electro-weak phase transition (with $T\sim 130$ GeV), the observed baryon asymmetry could be due to a PBH of $M\sim 10^{12}$ g and $a\approx 0.015$.

\section{Asymmetry without violating Hermiticity}

\begin{figure}
  \centering
  \includegraphics[width=0.45\textwidth]{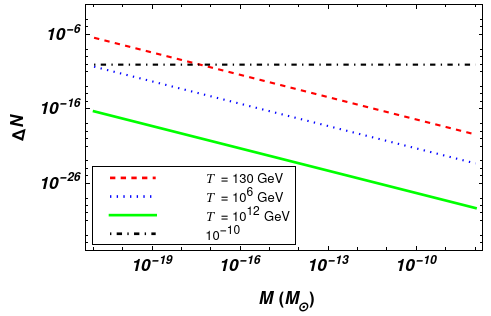}
  \caption{ Variation of $\Delta N$ of the Hermitian case at $r=2.1$ as a function BH mass for various  temperature of universe for a fixed spin $a=0.01$.}
  \label{herm}
\end{figure}
If we construct the total Lagrangian for neutrino and anti-neutrino together, then the anti-Hermitian vector term disappears. This is a good Lagrangian for the present purpose, when indeed we always consider neutrino and anti-neutrino together. This leads to the conservation of neutrino probability without the decay term in Eqs.\,(\ref{17})-(\ref{21}), which further suggests the observed $\Delta N$ to be at much smaller $a$. Fig.\,\ref{herm} shows the variation of $\Delta N$ in a range of PBH mass for different $T$ and fixed $a=0.01$. It confirms $\Delta N\sim 10^{-10}$ for $M=10^{-17}M_\odot$ and $T=130$ GeV. In fact at an earlier hotter universe with, e.g., $T=10^6$ GeV, the observed $\Delta N$ could be produced for $M=10^{-21}M_\odot$ at the same low spin. Note that for a fixed spin, the scaling of $\Delta N\sim 1/MT$. 

\section{Conclusion}
The origin of baryon asymmetry is an old problem in cosmology. While it has been attempted to uncover by many avenues, none of them appear to be foolproof, without caveats. One of the ideas is to generate baryogenesis through leptogenesis, what we rely upon. However, we have introduced a novel idea based on the weakly interacting neutrinos in early universe influenced by the spin-connection in the presence of gravitational field. We have shown that in general neutrinos interacting with BHs produce neutrino--anti-neutrino asymmetry, hence leptogenesis. Such an asymmetry in early universe produced by PBHs could be $10^{-10}$, depending on the mass and spin of BHs. From the preservation of $B-L$ symmetry and the violation of $B+L$ symmetry due to the sphaleron process, such a lepton asymmetry can lead to the equal baryon asymmetry settled by the electro-weak era, what we observe. The rate of sphaleron process gets exponentially suppressed below 130 GeV. Although it is possible to produce lepton asymmetry in our prescribed formalism below 130 GeV, it may not be viable to give rise to any further baryon asymmetry. Nevertheless, our model shows that different spins and masses of BH change the local asymmetry. The demonstrated asymmetry, particularly leptogenesis, is completely model independent, completely determined by the spacetime curvature coupling, which is an obvious effect in the early universe.

\section*{Acknowledgment}
The authors thank Nirmal Raj (IISc) for useful discussions, particularly for the underlying particle physics in early universe. Thanks are also due to Ranjan Laha (IISc) and Rajeev Jain (IISc) for discussions. MD acknowledges the financial support of the Ministry of Education (MoE) fellowship scheme. BM and AKJ thank the support from the project funded by SERB/ANRF, India, with Ref. No. CRG/2022/003460. MP acknowledges the Prime Minister’s Research Fellows (PMRF) scheme for providing fellowship.

\bibliography{reference}

\end{document}